\documentclass[11pt]{article}
\usepackage{amsmath, amssymb, amscd, amsthm, amsfonts}
\usepackage{graphicx}
\usepackage{hyperref}
\usepackage[affil-it]{authblk}
\usepackage[round,numbers]{natbib}
\usepackage[labelfont=bf, labelsep=period, justification=centering, skip=10pt]{caption}

\title{The Impact of Formations on Football Matches Using Double
Machine Learning. Is it worth parking the bus?}

\author[1]{Genís Ruiz-Menárguez}
\author[1]{Llorenç Badiella}
\affil[1]{Departament de Matemàtiques, Universitat Autònoma de Barcelona, Spain}

\date{}

\begin{document}

\maketitle

\begin{abstract}
This study addresses a central tactical dilemma for football coaches: whether to employ a defensive strategy, colloquially known as "parking the bus," or a more offensive one. Using an advanced Double Machine Learning (DML) framework, this project provides a robust and interpretable tool to estimate the causal impact of different formations on key match outcomes such as goal difference, possession, corners, and disciplinary actions. Leveraging a dataset of over 22,000 matches from top European leagues, formations were categorized into six representative types based on tactical structure and expert consultation. A major methodological contribution lies in the adaptation of DML to handle categorical treatments, specifically formation combinations, through a novel matrix-based residualization process—allowing for a detailed estimation of formation-versus-formation effects that can inform a coach's tactical decision-making.

Results show that while offensive formations like 4-3-3 and 4-2-3-1 offer modest statistical advantages in possession and corners, their impact on goals is limited. Furthermore, no evidence supports the idea that defensive formations—commonly associated with parking the bus—increase a team’s winning potential. Additionally, red cards appear unaffected by formation choice, suggesting other behavioral factors dominate. Although this approach does not fully capture all aspects of playing style or team strength, it provides a valuable framework for coaches to analyze tactical efficiency and sets a precedent for future research in sports analytics.
 
\end{abstract}
\textbf{Keywords:} Football, Formation, Double Machine Learning, Causal Inference, Confounding, XGBoost
\section{Introduction}\label{section-introduction}
Football coaches face a critical decision before each match: which formation to implement to maximize their team's performance. The organization of the eleven starting players into specific structural patterns is a foundational element of tactical strategy. This project aims to provide data-driven insights for this decision by analyzing whether certain defensive formations, commonly referred to as "parking the bus" \citep{guan2023parking}, are truly effective in specific scenarios. José Mourinho, former Chelsea FC manager, popularized this term to describe a rival's excessively defensive approach.

These defensive strategies are often employed against stronger opponents ~\citep{nazarudin2025evaluating,freitas2023elite}; yet their actual effectiveness remains a subject of debate. This study addresses this debate directly for the benefit of coaches and tactical analysts. We examine the direct relationship between formations and several key match outcomes, moving beyond anecdotal evidence to provide a robust statistical analysis. After generalizing and grouping formations based on tactical similarity and expert criteria, a Double Machine Learning framework is applied to estimate the isolated impact of each formation combination on various variables, such as goal difference. While formations may not be the sole predictor of match statistics, understanding whether specific systems consistently enhance or hinder key performance outcomes is invaluable for strategic planning.

Ultimately, the objective is to assess the isolated impact of formations themselves, providing coaches with a clear, unbiased look at how tactical systems, independent of individual team quality, influence match results.

\section{Data Processing}

Match data for this study were obtained from the Sportmonks API \citep{sportmonks2025}, a commercial provider of comprehensive football data. The dataset comprises over 22,000 professional league fixtures and includes a wide range of variables critical for our analysis, such as match statistics, team formations, and contextual information.

To ensure the dataset was robust and representative of modern football tactics, specific criteria were applied for data inclusion. The selection of seasons and leagues was guided by two primary considerations: regulatory consistency and tactical diversity.

\begin{enumerate}

\item \textbf{Season Selection:} The analysis includes seasons from 2018–2019 to 2024–2025. This broad temporal scope was chosen to ensure a sufficient data volume for robust statistical modeling. While the majority of this period reflects the modern tactical environment shaped by the "five substitutions" rule, seasons prior to this regulatory change (2018–2019, 2019–2020, and 2020–2021) were also included to significantly expand the dataset's size and statistical power. The effects of this regulatory change are implicitly captured by the inclusion of both league and season within the confounder set, which accounts for temporal and league-specific variations that may be correlated with the rule change.

\item \textbf{League Selection:} To create a geographically and tactically diverse dataset, matches from the seven highest-ranked European men's football leagues according to UEFA coefficients were included: the first divisions of England, Italy, Spain, Germany, France, the Netherlands, and Portugal. To further enhance the statistical power of the models, additional leagues were selected based on two criteria: a minimum of 32 regular-season matches and sufficient data availability from the API. This led to the inclusion of the top divisions of Turkey, Belgium, and Poland, as well as the second divisions of Spain, Italy, and England.

\end{enumerate}

\subsection{Confounding Variables}
The primary objective of this analysis is to isolate the intrinsic relationship between a team's formation and a rival's formation with respect to various match outcomes. To achieve this, a set of confounding variables has been selected. These variables are chosen to control for factors that influence both team strength and match outcomes but are not a direct result of the chosen formation. Consequently, mediator variables such as accumulated goals or possession, which are themselves outcomes of a team's playing style, will be intentionally excluded from the model.

The following variables are used as confounders:

\begin{itemize}
    \item \textbf{Match Context:} The season, league, or day of the week of the fixture.
    \item \textbf{Team Side:} A binary indicator for whether the main team played at home or away.
    \item \textbf{Team Strength Metrics:}
    \begin{itemize}
        \item The rate of accumulated points for both teams prior to the match.
        \item A binary flag indicating whether each team was simultaneously competing in the UEFA Champions League.
        \item The league ranking of each team at the time of the fixture.
        \item The winning streak from last matches.
    \end{itemize}
    \item \textbf{Environmental Factors:} Weather data for the match venue.
\end{itemize}

The final dataset included additional key variables for each fixture which are used as mediators, treatment variables or target variables at some point of the analysis:

\begin{itemize}

\item \textbf{Performance Metrics:} Goals, corners, possession, and disciplinary actions (red and yellow cards) for both teams.

\item \textbf{Tactical Information:} The formation system used by both teams (e.g., 4-3-3).

\item \textbf{Contextual Variables:} In-match events (e.g., penalties, substitutions), lineups, and weather data (e.g., temperature, humidity).

\end{itemize}

Variables that could act as mediators between a formation choice and the outcome—such as in-game substitutions or team fouls—were intentionally excluded from the set of confounders. This distinction is crucial to ensure that statistical model isolates the pure direct effect of the formation, rather than capturing indirect effects. While these variables are not used as confounders, some, such as possession and disciplinary actions, serve as secondary target variables for our analysis.

The raw dataset underwent several cleaning and filtering steps to ensure data quality and methodological consistency. Specific seasons from the Belgian Pro League and Turkish Süper Lig were excluded from the analysis due to incomplete match data, which would have compromised the integrity of feature engineering.

To focus on a consistent competitive environment, non-regular league fixtures, such as play-offs and play-outs, were removed from the dataset. Furthermore, to mitigate potential biases from early-season instability and late-season goal achievement effects, matches from the first two and final four rounds of each season were also discarded. This ensures that the analysis is based on matches where teams are at a comparable level of statistical performance and motivation.

To properly control for team quality and contextual factors, several variables were engineered to quantify team strength at the time of each fixture. These variables served as potential confounders in the relationship between treatment and outcome variables. Betting odds data were intentionally excluded from this set, as such information may implicitly incorporate factors related to formation choice, potentially introducing a source of endogeneity.

The engineered team strength variables include:
\begin{itemize}
    \item \textbf{Accumulated Points:} A numerical variable representing the ratio of points earned out of the total possible in prior matches, computed separately for home and away performances.
    \item \textbf{Champions League Flag:} A binary variable indicating whether the home or away team was competing in the UEFA Champions League at the time of the fixture.
    \item \textbf{Team Ranking:} An integer variable reflecting the league position of each team, with the same rank assigned to teams with equal points.
\end{itemize}

Consistent with the study's objective to isolate the pure causal effect of formations, variables that are structurally related to formations which are considered mediators -such as accumulated goals or fouls- were not included in the confounder set. This careful variable selection is crucial to prevent the model from capturing indirect effects and ensures that the estimated causal impact is a true reflection of formation choice itself.

\section{Methodology}

In order to estimate formations' impact on football match outcomes we propose to use a Double Machine Learning approach \cite{chernozhukov2018double}, that will remove the effect of confounders.

For the DML 1st step consider the following linear model:

\[
Y = D \beta + X \gamma + \varepsilon
\]

\begin{itemize}
    \item \( Y \): outcome variable (e.g., goal difference)
    \item \( D \): treatment variable (e.g., formation combination of both teams)
    \item \( X \): matrix of control variables (confounders),
    \item \( \beta\): causal parameter of interest,
    \item \( \gamma\): nuisance parameters,
    \item \( \varepsilon\): error term.
\end{itemize}

\subsection{Double Machine Learning}

Double Machine Learning states that the effect of a treatment variable $D$ on an outcome $Y$, after accounting for control variables $X$, can be obtained by first removing from both the treatment and the outcome the variation explained by the confounding variables, and then regressing the residuals of the outcome on the residuals of the treatment. The resulting coefficient corresponds to the partial effect $\beta$ of the treatment that is orthogonal to the confounding variables.

Function $\hat{f}(X)$ is defined as the Machine Learning estimator of \( Y \) on \( X \),  and \( \hat{g}(X) \) as the estimator of \( D \) on \( X \).

Both $\hat{f}(X)$ and $\hat{g}(X)$ can be adaptive to high dimensional spaces, and capture nonlinearities, interactions, and complex dependencies automatically. The final stage is obtained isolating the $\hat{\beta}$ estimator from the 3rd step residualization process.

Through sample splitting and cross-fitting, this modification maintains orthogonality and robustness while allowing the estimation of treatment effects in high-dimensional or nonlinear environments. Some examples of this extension include using boosting or decision trees models as a substitution of traditional linear models.

\subsection{New Approach for DML with categorical treatment}

Applying categorical treatment in the second DML step enables us to fully capture the partial effect of the target variable on each different treatment variable. This approach is highly effective when dealing with multicategorical and independent treatment variables.

To consider each formation combination in the variable $D$, dummy variables for $k^2-1$ unique formation combinations will be generated, where $k$ is the number of different formations and it is equal for both main and rival team.
The goal is to identify which formations are prone to contribute to a higher value in the target variable (e.g. goals difference), meaning winning by a larger margin in the case of goals. To do so, each treatment variable vector $D_{i,j}$ corresponds to the combination of the main team formation $i$ and rival team formation $j$.   
Therefore, the resulting models can be expressed in a two-dimensional matrix.

\[D_{i,j} = \alpha_{i,j} X + \varepsilon_{D_{i,j}}\]

Where,

\begin{itemize}
    \item $D_{i,j}$:  is a dummy vector which contains 1 for the combination of main team formation $i$ and rival team formation $j$ row, -1 for the unused formation row ($D_{k^2}$) and 0 otherwise.
    \item $\alpha_{i,j}$ is a vector of coefficients corresponding to each variable in the $X$ matrix. 
\end{itemize}

Note that the $D_{k,k}$ formation combination is ommitted, as it is not used to avoid multicollinearity. Nevertheless, this last formation data can be deduced from the other ones. 

This encoding scheme enables us to gather the true effect of each formation combination $D_{i,j}$ on the target variable $Y$. (See effect coding \cite{te2017size}).

\[r_{D_{i,j}} = D_{i,j} - \hat{D}_{i,j} \]

Where $r_{D_{i,j}}$ is the residual of the treatment variable $D_{i,j}$ prediction.

Consequently, the third model is computed:

\[
r_Y = \beta_{1,1} r_{D_{1,1}} + \cdots + \beta_{1,k} r_{D_{1,k}} + \cdots + \beta_{k,k-1} r_{D_{k,k-1}} + \epsilon
\]

The greater the $\beta_{i,j}$ is, the greater positive impact it will have on the target variable $Y$. 

The estimator $\hat{\beta}_{k,k}$ can be obtained as:

\[ \hat{\beta}_{k,k}
 = - (\hat{\beta}_{1,1}  + \cdots + \hat{\beta}_{1,k} + \cdots + \hat{\beta}_{k,k-1})
\]

Since $\beta_{k,k}$ belongs to the diagonal estimators, $\beta_{k,k}$ = 0.

\subsubsection{Categorical Treatment Properties}

By applying this new approach, the treatment $D$ is orthogonalized on the confounders $X$. After regressing each dummy $D_{i,j}$ on $X$ and taking residuals $r_{D_{i,j}}$, each residual $r_{D_i}$ is orthogonal to the controls $X$. As any correlation between $D_{i,j}$ and $Y$ that was due to $X$ has been purged, the remaining correlation between $r_{D_{i,j}}$ and $r_Y$ also captures the true causal effect. 
For every dummy variable $D_{i,j}$, a separate orthogonal moment condition is defined after residualization. Therefore, the orthogonality condition required by DML holds for every formation combination separately and the original framework properties remain valid. 

However, this new approach leads to a different interpretation of the parameter $\beta$. Each coefficient $\beta_{i,j}$ estimates the causal effect on goal difference of choosing the main team formation $i$ given the rival team formation $j$ in comparison to the omitted combination $D_{k,k}$.

\subsection{ Project Methodology}
To address the formation impact analysis, the standard DML framework was adapted. The core of our approach involves a two-stage residualization process, which can be summarized as follows:

\begin{enumerate}
    \item \textbf{First-Stage Residualization:} A machine learning model, specifically an XGBoost Regressor \cite{chen2016xgboost}, is used to predict the target variable (e.g., goal difference) based on the set of confounding variables. The resulting residual, denoted as $r_Y$, represents the component of the target variable not explained by these confounders.
    \item \textbf{Treatment Model Residualization:} Similarly, an XGBoost Regressor is trained to predict the treatment variable (formation combination) using the same set of confounders. To apply this to our categorical treatment, a separate regression model is computed for each formation combination. The resulting residuals, $r_D$, represent the variation in the treatment not explained by the confounders.
    \item \textbf{Causal Parameter Estimation:} Finally, the causal effect of formations is estimated by a simple linear regression of the target residuals ($r_Y$) on the treatment residuals ($r_D$). The coefficients from this final regression, $\hat{\beta}_{i,j}$, provide the unbiased causal estimates of the effect of each formation combination on the target variable.
\end{enumerate}

\section{Formation Analysis Results}
This approach consists of analyzing the impact of different tactical strategies on match outcomes. To do so, formations from the dataset were grouped into six distinct categories, sorted from most defensive to most offensive.

\subsection{Formation Treatment}
Formations are represented by a sequence of numbers, specifying the distribution of outfield players across the defensive, midfield, and offensive lines. Since the goalkeeper is not counted, the numbers sum to ten.

Initially, the dataset contained 28 distinct formations. To enable robust statistical analysis and mitigate the issues of data sparsity and lack of statistical power that would result from an excessively large treatment matrix, these formations were grouped into six representative categories. This process was conducted based on expert consultation and a tactical similarity criterion, ensuring that formations with a similar structural foundation were aggregated. For example, formations such as 4-3-3 and its variations (e.g., 4-3-1-2) were categorized together as they share a similar core structure. This grouping approach addresses the issue of different representations of the same underlying tactical system being treated as distinct entities.

The six resulting formations, sorted from the most defensive to the most offensive one, and their proportion in the dataset are shown in Table \ref{tab:formation_percentages_transposed}.

\vspace{-0.1cm}

\begin{table}[h!]
\centering
\caption{Formation percentages across different leagues}
\small
\setlength{\tabcolsep}{3pt}
\begin{tabular}{|l|c|c|c|c|c|c|}
\hline
\textbf{League} & \textbf{5-4-1} & \textbf{4-4-2} & \textbf{3-5-2} & \textbf{4-2-3-1} & \textbf{4-3-3} & \textbf{3-4-3} \\
\hline
\textbf{La Liga} & 8.5\% & 32.57\% & 10.79\% & 22.72\% & 20.64\% & 4.78\% \\
\textbf{Premier League} & 7.75\% & 15.32\% & 9.26\% & 31.23\% & 24.43\% & 12.01\% \\
\textbf{Serie A} & 1.38\% & 6.16\% & 33.24\% & 18.54\% & 25.45\% & 15.23\% \\
\textbf{Bundesliga} & 4.48\% & 13.9\% & 21.73\% & 26.68\% & 13.41\% & 19.81\% \\
\textbf{Ligue 1} & 8.72\% & 14.32\% & 15.65\% & 26.96\% & 20.17\% & 14.18\% \\
\hline
\end{tabular}
\label{tab:formation_percentages_transposed}
\end{table}

The most frequently used formations across all leagues are the 4-2-3-1 and 4-3-3. Furthermore, analysis confirmed that the proportion of formations used by home and away teams is nearly identical, which justifies a simplified modeling approach for the formation treatment.

\par
\vspace{0.2cm}
Find below some fixture statistics on average grouped by formation and still ordered from most defensive to most offensive ones.

\begin{table}[h!]
    \centering
    \caption{Average Match Statistics by Formation across All Leagues}
    \label{tab:average_stats_by_formation}
    \small
    \setlength{\tabcolsep}{3.5pt}
    \begin{tabular}{|c|c|c|c|c|c|c|}
        \hline
        \textbf{Formation} & \textbf{5-4-1} & \textbf{4-4-2} & \textbf{3-5-2} & \textbf{4-2-3-1} & \textbf{4-3-3} & \textbf{3-4-3} \\
        \hline
        \textbf{Goals} & 1.236 & 1.249 & 1.268 & 1.392 & 1.437 & 1.341 \\
        \textbf{Red Cards} & 0.110 & 0.114 & 0.104 & 0.106 & 0.103 & 0.098 \\
        \textbf{Yellow Cards} & 2.169 & 2.246 & 2.215 & 2.110 & 2.075 & 2.188 \\
        \textbf{Possession} & 48.98\% & 48.62\% & 48.24\% & 50.72\% & 51.63\% & 49.80\% \\
        \textbf{Corners} & 4.751 & 4.719 & 4.754 & 5.004 & 5.107 & 4.854 \\
        \hline
    \end{tabular}
\end{table}
\par 

Although some formations, such as the 4-2-3-1 and 4-3-3, are associated with higher average goals and possession, these preliminary correlations are not sufficient for a valid causal inference. The observed statistical relationships are likely a result of confounding factors and complex interactions. The nominal formation used is a simplification of a team's dynamic tactical structure, which can change multiple times during a match due to in-game adaptations, substitutions, and opponent responses. Furthermore, a single formation can accommodate a wide variety of playing styles, which are not captured by a simple numerical representation. Therefore, it is crucial to employ a robust causal inference framework to isolate the true effect of a formation from these confounding variables.

\subsection{Model Performance}
Five separate XGBoost Regressor models were trained, each corresponding to a different target variable. The hyperparameter tuning for each model was performed aiming to maximize the negative mean squared error (-MSE). The MSE is the primary metric for evaluating the predictive performance of the first-stage regressions, as a value of zero corresponds to a perfect fit.

A critical hyperparameter choice was setting the upper limit for \textit{max\_depth} to 5. This constraint limits the complexity of each individual tree, which reduces the model's susceptibility to overfitting and helps ensure a lower variance in the predictions. Consequently, this approach also helps maintain symmetry in the final $\hat{\beta}$ estimations by preventing the model from making overly specific approximations.

The predictive performance of each model, evaluated on a held-out test set, is summarized in the table below. The $R^2$ metric is provided to quantify the proportion of variance in each target variable explained by the set of confounders.

\newpage

\begin{table}[h!]
    \centering
    \caption{Predictive Performance Metrics for First-Stage XGBoost Models}
    \label{tab:model_performance}
    \small
    \begin{tabular}{|l|c|c|}
        \hline
        \textbf{Statistic} & \textbf{MSE} & \textbf{R$^2$} \\
        \hline
        Goals & 2.70 & 0.118 \\
        \hline
        Red Cards & 0.19 & 0.003 \\
        \hline
        Yellow Cards & 3.18 & 0.030 \\
        \hline
        Possession & 343.35 & 0.298 \\
        \hline
        Corners & 7.98 & 0.781 \\
        \hline
    \end{tabular}
\end{table}

As the table shows, the mean squared error values are not directly comparable across models due to the differing units and scales of the target variables (e.g., possession is a percentage, while red cards are discrete counts). The model predicting corners exhibits the highest $R^2$ value, indicating it captures the largest proportion of variance among the five targets.

It is important to note that maximizing predictive accuracy was not the primary goal of this study. The first-stage models serve as a means to an end, specifically to perform the residualization necessary for obtaining the DML-estimated $\hat{\beta}_{i,j}$ coefficients. Given the high degree of unpredictability inherent in football, $R^2$ values for many match outcomes are expected to be low.

\subsection{Formation Matrix}

This DML approach enabled the computation of several causal analyses regarding the impact of formation combinations on various match outcomes. To interpret the resulting matrices of estimators, it is important to recall some key theoretical properties of the framework:
\begin{itemize}
    \item The diagonal of the matrix represents a formation's effect against itself, and thus, its value is defined as zero.
    \item The matrix is theoretically symmetrical, with $\hat{\beta}_{i,j} = -\hat{\beta}_{j,i}$. Any minor deviations from perfect symmetry are attributed to the approximations made by the machine learning models.
\end{itemize}

The statistical significance of the estimators is determined by p-values, $\{^{***}p < 0.001, \ ^{**}p < 0.01, \ ^{*}p < 0.05, \ \text{ns } p >= 0.05\}$. Cells with a p-value below 0.05 are considered statistically significant, indicating a rejection of the null hypothesis that the $\hat{\beta}_{i,j}$ estimator is equal to zero.

The following matrices present the pure causal effect between formation combinations for a specific target variable. To provide a more complete tactical context, these estimations can be adjusted for the effects of playing home or away. The average effect of the target variable for a home team is estimated and added to the causal coefficient, providing a close approximation of the total side effect.

This calculation is represented as:
\[
\hat{\beta}_{side_{i,j}} = \hat{\beta}_{i,j} + E(\hat{Y}_{home})
\]
where $E(\hat{Y}_{home})$ is the estimated average effect on the target variable for the home team (e.g., goals difference).

\subsubsection{Target Variable 1: Goals Difference}
The resulting $\hat{\beta}_{i,j}$ estimators for goals difference as the target variable are:

\vspace{-0.1cm}
\begin{figure}[h!]
    \centering
    \includegraphics[width=0.75\textwidth]{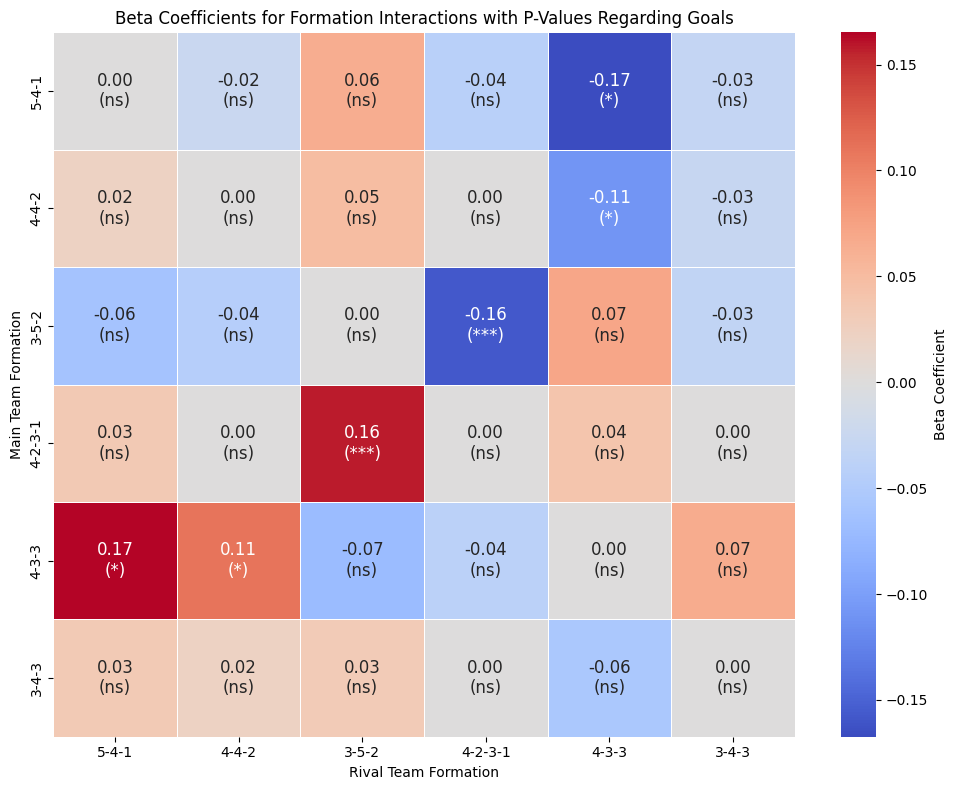}
    \caption{$\hat{\beta}_{i,j}$ coefficients and p-values for goals difference}
    \label{fig:betas_goals_diff}
\end{figure}
\vspace{-0.1cm}
The analysis of the estimators reveals three statistically significant $\hat{\beta}_{i,j}$ coefficients, each with a p-value below the 0.05 threshold. These significant estimators are:

\begin{itemize}
    \item \textbf{4-2-3-1 vs. 3-5-2:} This combination provides an estimated advantage of 0.16 goals for the 4-2-3-1 formation ($p < 0.001$), suggesting strong statistical significance.
    \item \textbf{4-3-3 vs. 5-4-1:} This combination shows an advantage of 0.17 goals for the 4-3-3 formation ($p < 0.05$).
    \item \textbf{4-3-3 vs. 4-4-2:} This combination results in a 0.11 goal advantage for the 4-3-3 formation ($p < 0.05$).
\end{itemize}

All other estimated coefficients for goal difference are not statistically significant and, thus, can be considered zero.

To provide a more comprehensive tactical context, the pure formation effect can be combined with the well-established home-field advantage. The average home-team effect, denoted as $E(\hat{Y}_{\text{home}})$, is estimated to be \textbf{0.285} goals when the target variable is goals difference. Consequently, the total estimated advantage of a 4-2-3-1 formation playing at home against a 3-5-2 would be approximately \textbf{+0.445} goals ($0.16 + 0.285$).

The finding that only a small number of estimators are significant suggests that formation choice alone does not have a major isolated impact on goal difference.

\subsubsection{Target variable 2: Red Cards difference}
The DML model was also applied to the difference in red cards between the main and rival teams, considering both direct red cards and those resulting from a second yellow card. In this analysis, the data were not stratified by home or away team, as the goal was to assess the pure, unadjusted causal effect of each formation combination on the number of red cards.

% \vspace{-0.5cm}
\begin{figure}[h!]
    \centering
    \includegraphics[width=0.75\textwidth]{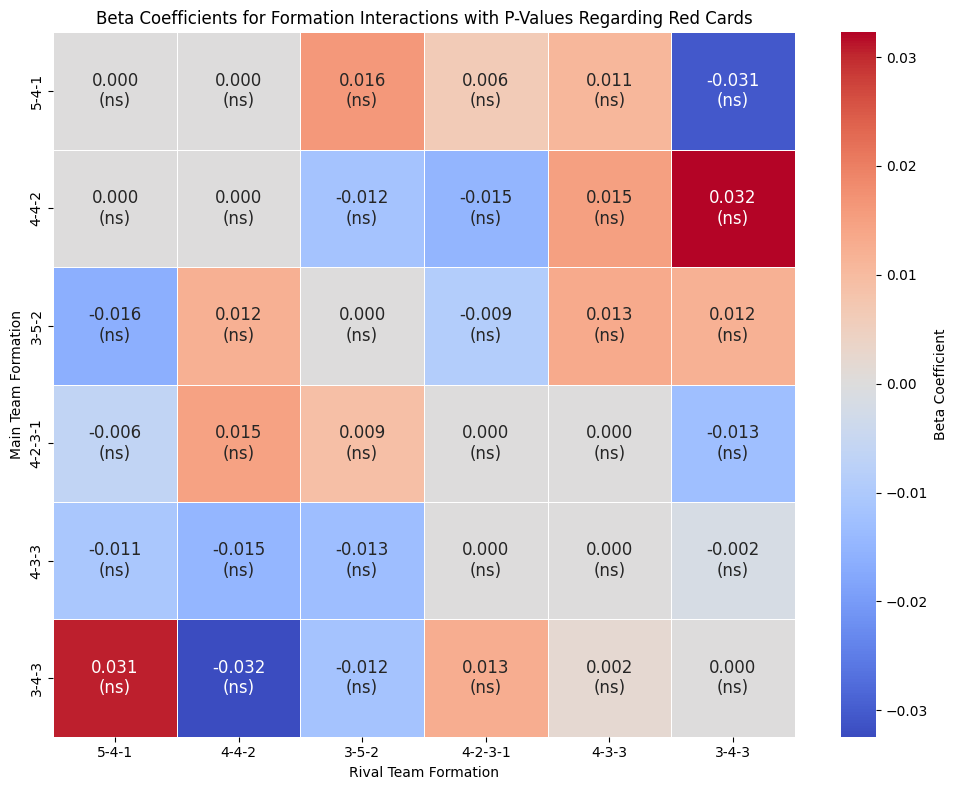}
    \caption{$\hat{\beta}_{i,j}$ coefficients and p-values for red cards difference}
    \label{fig:betas_red_cards_diff}
\end{figure}
\vspace{-0.6cm}

The results indicate that there are no statistically significant estimators for the red cards difference, as all p-values are greater than the 0.05 threshold. Consequently, we cannot reject the null hypothesis, and the estimated coefficients can be considered zero ($\hat{\beta}_{i,j} = 0 \quad \forall i,j$).

This lack of significance suggests that red cards are not causally influenced by the specific formation combinations, particularly after controlling for confounding factors such as team strength and general match conditions. This finding aligns with the understanding that red cards are often a result of random, unpredictable match events and are more closely linked to a player's individual intensity and behavior, which are difficult to quantify.

\subsubsection{Target Variable 3: Yellow Cards Difference}
The DML model was also applied to the difference in yellow cards between the main and rival teams, with the goal of assessing the pure causal effect of each formation combination on this outcome. This analysis did not stratify the data by home or away team.

\begin{figure}[h!]
    \centering
    \includegraphics[width=0.75\textwidth]{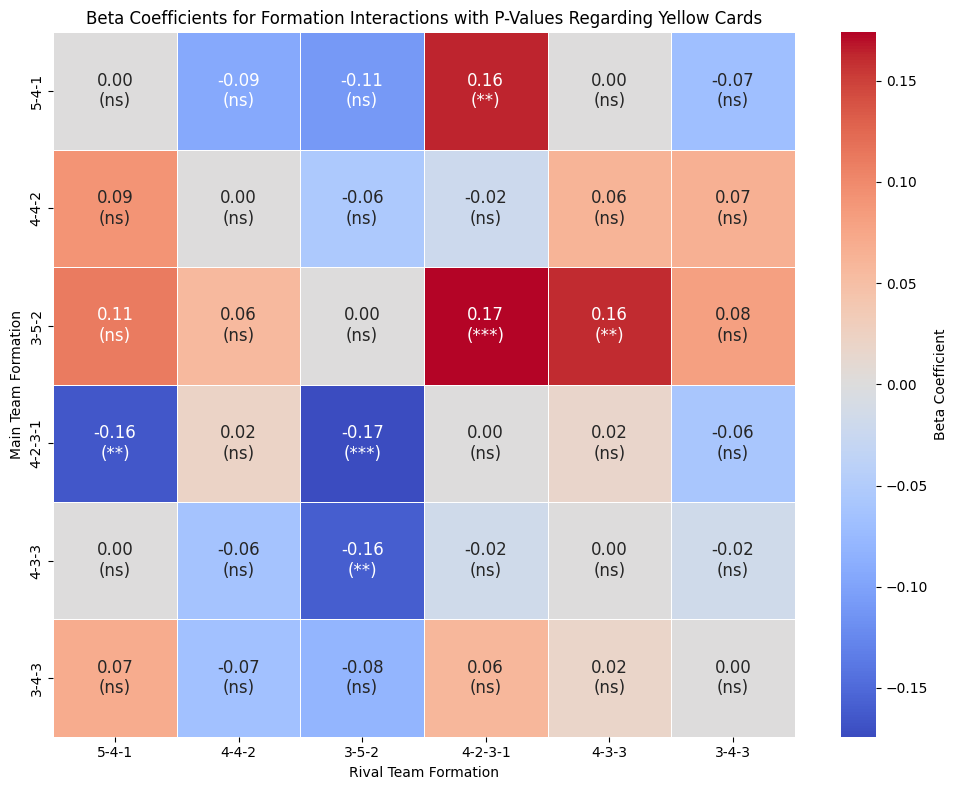}
    \caption{$\hat{\beta}_{i,j}$ coefficients and p-values for yellow cards difference}
    \label{fig:betas_yellow_cards_diff}
\end{figure}

The results reveal three statistically significant pairs of coefficients. As the matrix is symmetrical ($\hat{\beta}_{i,j} = -\hat{\beta}_{j,i}$), these represent three unique formation combinations:

\begin{itemize}
    \item \textbf{5-4-1 vs. 4-2-3-1:} The 5-4-1 formation is associated with an estimated increase of 0.16 yellow cards on average ($p < 0.01$). This suggests that a team in a 5-4-1 formation would receive approximately one additional yellow card every six matches when facing a team in a 4-2-3-1 formation.
    \item \textbf{3-5-2 vs. 4-2-3-1:} The 3-5-2 formation is associated with a larger estimated increase of 0.17 yellow cards ($p < 0.001$), indicating a strong statistical significance.
    \item \textbf{3-5-2 vs. 4-3-3:} This combination results in a 0.16 yellow card increase for the 3-5-2 formation ($p < 0.01$).
\end{itemize}

These findings suggest that defensive formations incur a subtle increase in yellow card penalties when compared to more offensive ones. However, as noted by Badiella Busquets \cite{badiella2023influence}, disciplinary actions such as yellow cards are also heavily influenced by individual player intensity, which is not captured by these models.

\subsubsection{Target variable 4: Possession}
The DML model was also applied to analyze the causal impact of formation combinations on possession. The hypothesis is that a team's tactical structure significantly influences its ability to control the ball during a match.

\begin{figure}[h!]
    \centering
    \includegraphics[width=0.75\textwidth]{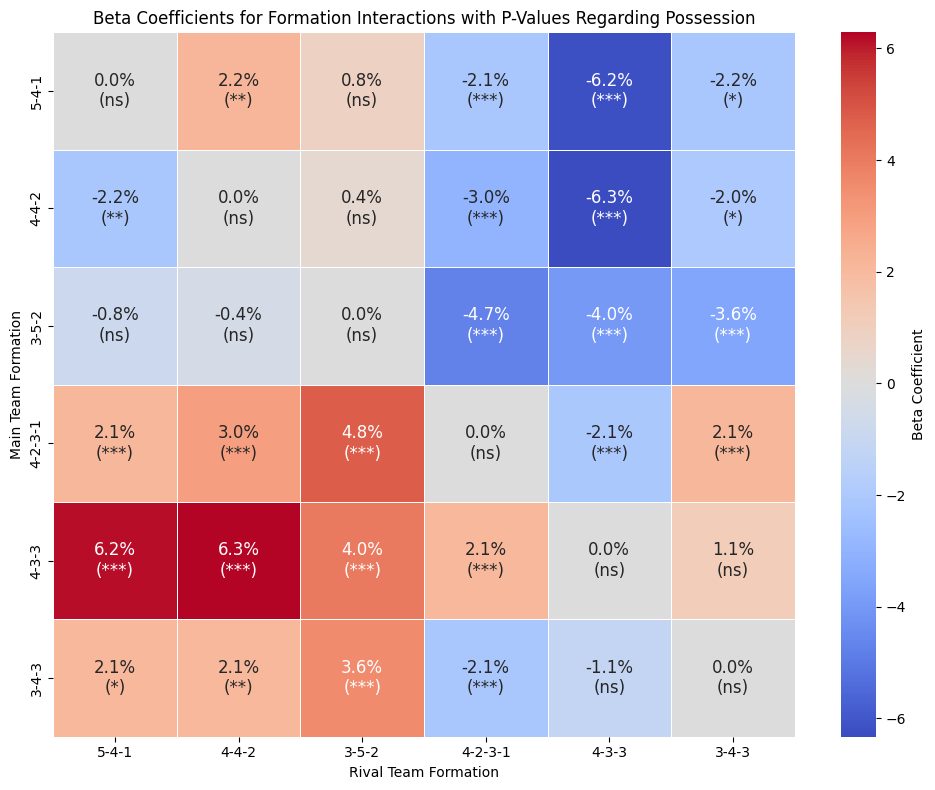}
    \caption{$\hat{\beta}_{i,j}$ coefficients and p-values for possession difference}
    \label{fig:betas_possession_diff}
\end{figure}

% \newpage

The results provide valuable insights into the intrinsic behavior of formations with respect to possession. The matrix reveals that 13 of the 18 unique formation combinations show statistically significant coefficients, with most having a p-value of $<0.001$.

The matrix can be divided into two key regions: a top-right blue region and a bottom-left red region, which represent offensive-defensive matchups.
\begin{itemize}
    \item \textbf{Offensive vs. Defensive Matchups (Top-right):} This region clearly shows that defensive formations tend to have less control over possession. The estimated possession difference between defensive formations (5-4-1, 4-4-2, 3-5-2) and offensive ones (4-2-3-1, 4-3-3, 3-4-3) is consistently significant, with all p-values less than 0.001. This indicates that formations with more midfielders and fewer defenders are more effective at retaining possession.
    \vspace{-0.1cm}
    \item \textbf{Defensive vs. Offensive Matchups (Bottom-left):} This region is a symmetrical reflection of the top-right, confirming that offensive formations have greater ball control when facing defensive ones.
\end{itemize}

\vspace{-0.1cm}

Overall, the 4-3-3 formation demonstrates the strongest positive effect on ball control when compared to other offensive formations. Interestingly, the 3-4-3, despite being highly offensive, shows less ball dominance than the 4-2-3-1, which could be due to a more direct, vertical playing style with fewer players in midfield to calmly circulate the ball.
\vspace{-0.1cm}

\subsubsection{Target variable 5: Corners}
Finally, the influence of formations on corner kicks was also analyzed. The hypothesis is that formations with higher possession and a more offensive structure will generate more scoring opportunities, consequently leading to a higher number of corners.

\vspace{-0.1cm}

\begin{figure}[h!]
    \centering
    \includegraphics[width=0.75\textwidth]{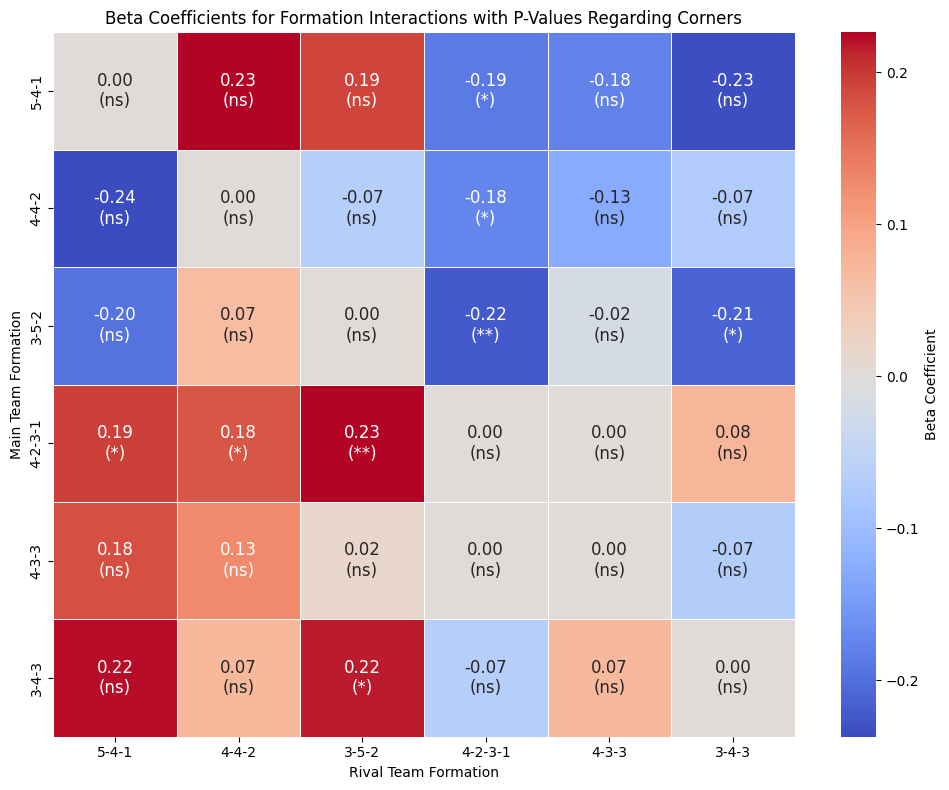}
    \caption{$\hat{\beta}_{i,j}$ coefficients and p-values for corners difference}
    \label{fig:betas_corners_diff}
\end{figure}

The matrix of coefficients reveals that offensive formations have a statistically significant positive effect on corner kicks when facing defensive ones. This aligns with the findings from the possession analysis, as teams with more ball control and a more aggressive tactical setup create more attacking opportunities. For example, the 4-2-3-1, the most common formation in our dataset, shows a positive estimated effect on corners when paired against defensive systems like the 5-4-1 (+0.19), 4-4-2 (+0.18), and 3-5-2 (+0.23). While these effects are statistically significant, their practical magnitude is small. Given that the average number of corners per team is around five per match, the estimated changes are not substantial enough to be considered a major tactical advantage.
 
\section{Discussion}

This study reveals that while football formations provide a baseline tactical framework, they are not sufficient to define or predict a team’s playing style or match outcome. Each nominal formation—such as 4-3-3 or 4-2-3-1 can embody a variety of tactical behaviors, and the assumption that the starting formation remains static throughout the match overlooks the dynamic nature of in-game adaptations. Furthermore, some teams employ different formations when attacking compared to when they are defending.

Through the application of Double Machine Learning—redefined in this project to handle the categorical treatment through a novel matrix-based residualization of formation combinations—it becomes evident that defensive formations, popularly known as "parking the bus", generally serve to neutralize offensive threats rather than produce statistically advantageous outcomes in terms of goals, possession, corners, or disciplinary measures. However, "parking the bus" often implies other strategies not considered in this analysis, such as time-wasting, tactical fouls, and slowing down the pace of the game.

Conversely, offensive formations like 4-2-3-1, 4-3-3, and 3-4-3 show a modest but statistically significant positive impact on possession and corner differences when facing more defensive systems. However, formation choice alone has minimal impact on red cards, which appear to be driven by more random or behavioral factors. Importantly, the causal effect of formations on goal difference is limited, with only three formation combinations yielding significant estimators—underscoring that formations do not fully account for scoring dynamics. Therefore, there is no statistical evidence that "parking the bus" offers any scoring or winning advantage.

While these results provide insightful patterns, they come with certain limitations, including the criteria used for defining team strength and the simplification involved in formation grouping, which may not fully explain all aspects of playing behavior. Nonetheless, this serves as a valuable starting point for deeper formation-based performance analysis and provides a clear, data-driven framework that can help coaches make more informed tactical decisions.

\section{Acknowledgements}

We would like to express our sincere gratitude to Juan Jesús Rodríguez and Juan Camilo Vázquez for their invaluable support throughout this project. Their deep football knowledge, especially in helping group formations meaningfully and providing tactical insights, was essential to interpreting the data accurately and grounding the analysis in real-world football logic. 

\section{Author contributions}
GR and LB conceived the study and outlined the article content. GR constructed the dataset, performed the statistical analyses and wrote the first draft of the article. GR and LB contributed in the methodological approach and reviewed of the manuscript. All authors have read and approved the final version of the article.

\section{Funding}
This work was partially funded by the grant RTI2018-096072-B-I00 from the Spanish Ministry of Science, Innovation and Universities

\bibliographystyle{unsrt}
\bibliography{references}

\end{document}